\documentclass[preprint,prd,showpacs,superscriptaddress]{revtex4}
\usepackage{amssymb,amsmath,graphicx,amscd}

\def\E{\mathbf{E}}
\def\B{\mathbf{B}}
\def\D{\mathbf{D}}
\def\H{\mathbf{H}}
\def\P{\mathbf{P}}
\def\x{\mathbf{x}}
\def\k{\mathbf{k}}

\begin{document}

\title{An Analog Model for Quantum Lightcone Fluctuations in Nonlinear Optics }

\author{L. H. Ford}
\email{ford@cosmos.phy.tufts.edu}
\affiliation{Institute of Cosmology, Department of Physics and Astronomy,
Tufts University, Medford, Massachusetts 02155, USA}

\author{V. A. De Lorenci}
\email{delorenci@unifei.edu.br}
\affiliation{Instituto de Ci\^encias Exatas, Universidade Federal de Itajub\'a, Itajub\'a, MG 37500-903, 
Brazil}

\author{G. Menezes}
\email{gsm@ift.unesp.br}
\affiliation{Instituto de F\'\i sica Te\'orica, Universidade Estadual Paulista,\\
Rua Dr. Bento Teobaldo Ferraz 271 - Bloco II, 01140-070  S\~ao Paulo, SP, Brazil}

\author{N. F. Svaiter}
\email{nfuxsvai@cbpf.br}
\affiliation{Centro Brasileiro de Pesquisas F\'{\i}sicas, Rua Dr. Xavier
Sigaud 150, 22290-180  Rio de Janeiro, RJ, Brazil}

\begin{abstract}
We propose an analog model for quantum gravity effects using nonlinear dielectrics. 
Fluctuations of the spacetime lightcone are expected in quantum gravity, leading to
variations in the flight times of pulses. This effect can also arise in a nonlinear material.
We propose a model in which fluctuations of a background electric field, such as that
produced by a squeezed photon state, can cause fluctuations in the effective lightcone
for probe pulses. This leads to a variation in flight times analogous to that in quantum
gravity. We make some numerical estimates which suggest that the effect might be
large enough to be observable.
\end{abstract}


\maketitle

\baselineskip=14pt

\section{Introduction}
\label{sec:intro}

A quantum theory of gravity is expected to predict quantum fluctuations of the spacetime
geometry, and hence of the lightcone. This effect was recognized long ago by early
workers on quantum gravity~\cite{Pauli,Deser,DeWitt}. In particular, Pauli~\cite{Pauli}
 once suggested that lightcone fluctuations
might remove the ultraviolet divergences of quantum field theory by smearing the 
singular behavior of Green's functions. This hope has not yet been realized, but still remains
a possibility, given that a complete quantum theory of gravity has not yet been found.
Several proposals for the small scale structure of spacetime near the Planck scale have been
made, including that of ``spacetime foam"~\cite{Wheeler,Hawking}. For a recent
discussion of work on this topic, including observational bounds, see Ref.~\cite{Ng}. 
The possibility of an energy-dependent refractive index in one model was discussed
in Ref.~\cite{EMN00}.
Even in the absence of a full theory, it is possible to discuss lightcone fluctuations in weak
field quantum gravity. When linearized perturbations of a fixed background are quantized,
it becomes possible to treat smearing of Green's functions and the associated lightcone
fluctuations~\cite{F95,FS96,FS97}. 
These are the {\it active} fluctuations, coming from the dynamical degrees
of freedom of gravity itself. There are also {\it passive} fluctuations, which are driven by
the fluctuations of the stress tensor of matter fields~\cite{FW08}. 

One expects quantum gravity effects to be very small, except in extreme conditions
such as the early universe or near small black holes. For this reason, it is of interest
to seek analog models in condensed matter systems, where similar effects might be
exhibited but be much larger. The use of analog models for black hole evaporation
has been a very active area of research, and is reviewed extensively in Ref.~\cite{BLV}. 
The effects of spatial randomness in a material have recently been treated as an analog 
model for lightcone fluctuations~\cite{KMN10,AGKMS11}. In the present
paper, we wish to propose a different model based on nonlinear optics, where fluctuations
of the effective dielectric function lead to a form of lightcone fluctuations.

There has long been interest in the parallels  between light propagation in gravitational
fields and in dielectric media. Gordon~\cite{Gordon} showed that a gravitational field
may be used to mimic the effects of a dielectric. Conversely, it is possible to define a dielectric
medium which mimics the effect of a gravitational field~\cite{Moller}. In the latter case,
the medium must have the property that the magnetic permeability is equal in magnitude
to the dielectric permittivity in Gaussian units. This is not realized by any known material.
Our interest is not in attempting to reproduce the detailed effects of gravity in a medium,
but rather in the fact that a realistic material alters the speed of light in the material. When
this speed can fluctuate, a form of lightcone fluctuations arises. 

The outline of the paper is as follows: In Sect.~\ref{sec:gravity}, we review selected
aspects of lightcone fluctuations in linearized quantum gravity. The crucial results from nonlinear 
optics which are needed will be discussed in Sect.~\ref{sec:wave}. In Sect.~\ref{sec:flucts}
we introduce our analog model, and derive a key result for the variance in photon flight
times in the presence of the lightcone fluctuations. Some numerical estimates are given
in Sect.~\ref{sec:est}, where we discuss the possibility of a laboratory experiment to 
look for the lightcone fluctuation effect. Our results are summarized and discussed in
Sect.~\ref{sec:sum}. We will employ SI units, except in  Sect.~\ref{sec:gravity}, where
$\hbar = c =1$ units are used.

\section{Lightcone Fluctuations in Linearized Quantum Gravity}
\label{sec:gravity}

In this section, we review selected aspects of lightcone fluctuations in quantum gravity.
First, let us recall the situation in classical general relativity theory, where the lightcone
is a dynamical object determined by the spacetime geometry. A small perturbation of the
geometry alters the lightcone.  If we start with a background of Minkowski spacetime,
and add small perturbations, the apparent speed of light as measured by the flat background
metric can either increase or decrease. In the case of  a weak field Schwarzschild metric,
the change is a decrease, leading to the well-known Shapiro time delay~\cite{Shapiro}. 
The measurement of the time delay of radar signals passing near the sun is one of the 
better experimental tests of general relativity~\cite{Will}. 
Similarly, a Schwarzschild metric with a negative mass would
lead to a time advance, or an apparent increase in the speed of light compared to flat
spacetime. Of course, the local speed of light is constant; it is the apparent speed measured
over a finite distance which changes.

Now suppose that the linearized perturbations are quantized and are subject to quantum
fluctuations. This will lead to fluctuations in the apparent speed of light and hence of the lightcone.
Let the metric be written as
\begin{equation}
g_{\mu\nu} = \eta_{\mu\nu} + h_{\mu\nu}\,,
\end{equation}
where $\eta_{\mu\nu}$ is the Minkowski metric, and $h_{\mu\nu}$ is the quantized linear 
perturbation, which now becomes the graviton field operator. Let $\sigma(x,x')$ be the
invariant interval function, defined as one-half of the squared geodesic distance between
spacetime points $x$ and $x'$. In flat spacetime, $\sigma = \sigma_0$, where 
\begin{equation}
\sigma_0 = \frac{1}{2} (x-x')^2 \,.
\end{equation}
Expand $\sigma$ in powers of $h_{\mu\nu}$ as
\begin{equation}
\sigma = \sigma_0 + \sigma_1 + \sigma_2 + \cdots \,,
\end{equation}
so that $\sigma_1$ is first order, $\sigma_2$ is second order, {\it etc}.
In flat spacetime, the retarded Green's function for a massless field is a delta function
on the forward lightcone:
\begin{equation}
G_0(x,x') = \frac{\theta(t-t')}{4 \pi} \, \delta(\sigma_0)\,.   \label{eq:G0}
\end{equation}
In a fixed curved spacetime, the form of the retarded Green's function near the lightcone
has the same functional form, but with $\sigma_0$ replaced by $\sigma$. If we now suppose 
that  $h_{\mu\nu}$,  and hence $ \sigma_1$, undergo Gaussian fluctuations around a mean
value of zero,   then the averaged Green's function becomes a Gaussian centered on the 
mean lightcone~\cite{F95}:
\begin{equation}
\langle G(x,x') \rangle = \frac{\theta(t-t')}{8 \pi^2} \, \sqrt{\frac{2\pi}{\langle  \sigma_1^2 \rangle}}
\, \exp \left( -\frac{\sigma_0^2}{2\langle \sigma_1^2 \rangle} \right)\,. \label{eq:G-av1}
\end{equation}
In the limit that $\langle  \sigma_1^2 \rangle \rightarrow 0$, the Gaussian in 
Eq.~(\ref{eq:G-av1}) approaches a delta function, and the averaged Green's
function approaches the form in Eq.~(\ref{eq:G0}).  In this case, the mean lightcone is
that of Minkowski spacetime. If we include the second order term,  $\sigma_2$, then
Eq.~(\ref{eq:G-av1}) is modified to become
\begin{equation}
\langle G(x,x') \rangle = \frac{\theta(t-t')}{8 \pi^2} \, \sqrt{\frac{2\pi}{\langle  \sigma_1^2 \rangle}}
\, \exp \left[ -\frac{(\sigma_0 + \langle \sigma_2 \rangle)^2}
{2\langle \sigma_1^2 \rangle} \right]\,. 
\label{eq:G-av2}
\end{equation}
We notice that these two last equations appear misprinted in Ref. \cite{F95}.
The effect of the $\sigma_2$ contribution is simply to shift the mean lightcone from that
of flat spacetime to a curved spacetime lightcone. One way to create lighcone fluctuations
is with a bath of gravitons in a non-classical state, such as a squeezed vacuum state~\cite{F95}.
In this case, $\langle  \sigma_1 \rangle =0$, so to lowest order, the lightcone fluctuates
around that of flat spacetime.
The shift of the mean lightcone described by $\langle \sigma_2 \rangle$  reflects the curving 
of the background spacetime by the gravitons. 

Here $\langle  \sigma_1^2 \rangle$ and  $\langle  \sigma_2 \rangle$  are understood 
to be finite expectation values. If we are interested in the effect of gravitons in a non-classical 
state, then we may take the difference between the formal expectation values in the
state in question, and in the Minkowski vacuum state. More generally, finite
 expectation values can arise from averaging two-point functions over finite regions
 of spacetime.

The replacement of a delta function by a Gaussian is the mathematical expression of
lightcone fluctuations. The physical consequence  of the smearing of the lightcone is
that the speed of propagation of pulses between a source and a detector becomes
statistical.  Some pulses will travel slower than the mean speed, as measured in the background spacetime, but others travel faster, with
a Gaussian distribution of flight times centered on the classical flight time in flat
spacetime. This leads  to a mean variation in flight times of   
\begin{equation}
\Delta t = \frac{\sqrt{\langle  \sigma_1^2 \rangle}}{r} \, , 
\end{equation}
where $r$ is the distance (measured in the flat background) between the source and
the detector. This variation may be expressed in terms of the graviton two-point 
function $\langle h_{\mu\nu}(x) h_{\alpha\beta}(x') \rangle$ as
\begin{equation}
(\Delta t)^2 = \frac{1}{4} \, \int_0^r dt \,dt'  \, n^\mu n^\nu n^\alpha n^\beta \,
\langle h_{\mu\nu}(x) h_{\alpha\beta}(x') \rangle \,. \label{eq:dt2-grav}
\end{equation}
Here $n^\mu$ is a unit spacelike vector defining the direction between the source
and detector, and the integrations are along the mean lightcone. This expression
may be shown to be gauge invariant provided that the metric perturbation $h_{\mu\nu}$
is localized between the source and the detector~\cite{YF99}. 
Equation~\eqref{eq:dt2-grav}
may also be derived from the Riemann tensor correlation function~\cite{YSF09}.
Integrals of the Riemann tensor correlation function may also be used to obtain other
physical effects of spacetime geometry fluctuations, such as luminosity 
variations~\cite{BF04} and spectral line broadening and angular blurring~\cite{TF06}.

 It is important to note that $\Delta t$ is the ensemble averaged variation in flight times,
 not necessarily the expected variation in flight times of two successive pulses~\cite{FS96}. If the
 two pulses are sufficiently close in time, they are correlated and tend to have flight times
 which differ by less than $\Delta t$. This can be understood as correlated pulses
 probing approximately the same classical spacetime geometry. A given source of
 metric fluctuations, such as a bath of gravitons, will define a characteristic fluctuation
 time scale of $\tau$. Pulses emitted at time intervals less than about $\tau$ are
 correlated, whereas those emitted with longer separations become uncorrelated and have
 flight time differences of order $\Delta t$.  If the metric fluctuations are due to a bath
 of gravitons with characteristic wavelength $\lambda_g$, then pulses are uncorrelated
 when their times of emission differ by more than about  $\lambda_g$. 
 
 For the case of a squeezed vacuum state of gravitons, pulses traveling in the same direction as
 the gravitons have a flight time variation of the order of 
 \begin{equation}
\Delta t \approx \ell_p \, \lambda_g\,\sqrt{U} \, r \,,   \label{eq:sq-grav}
\end{equation}
where $\ell_p$ is the Planck length, and $U$ is the characteristic energy density of the
gravitons. Clearly, this is a very small effect in the present 
 day universe. The presence of the Planck length in Eq.~(\ref{eq:sq-grav}) can be traced
 to the fact that the graviton two-point function is proportional to $\ell_p^2$, and is hence a 
 universal feature of lightcone fluctuations in quantum gravity. A major motivation for
analog models is to find systems with larger effects.

\section{ Wave propagation in nonlinear optics}
\label{sec:wave}

\subsection{Wave Equations}
\label{sec:wave_eq}

Unlike general relativity, Maxwell's theory is fundamentally a linear theory. However,
interactions of electromagnetic fields with charges can produce an effective nonlinear
wave equation. In nonlinear optics, this arises because atoms behave as nonlinear
oscillators in strong electric fields. For a review, see for example, Ref.~\cite{Boyd}.
In the absence of free charges or currents, electromagnetic waves in a nonlinear material 
medium is described by the usual Maxwell equations:
\begin{equation}
\nabla \cdot \B = 0 \,, \quad \nabla\times \E = - \frac{\partial \B}{\partial t}\,, \quad
\nabla\cdot \D = 0\,, \quad \nabla\times \H =  \frac{\partial \D}{\partial t} \,,
\label{eq:Maxwell}
\end{equation}
together with the constitutive relations for  a nonmagnetic, but polarizable material,
\begin{equation}
\B = \mu_0 \H\,, \qquad   \D = \epsilon_0\E + \P \,.
\label{eq:contitutive}
\end{equation}
Here $\E$ and $\B$ stand for the fundamental electric and magnetic fields, respectively,
while $\D$ and $\H$ are the corresponding induced fields. The relation between 
the polarization vector $\P$ and $\E$ is nonlinear, and can be expanded
in a power series as
\begin{equation}
P_i = \epsilon_0 \left(\chi_{ij}^{(1)} E_{j} + \chi_{ijk}^{(2)} E_{j}E_{k} 
+ \chi_{ijkl}^{(3)} E_{j}E_{k}E_{l} + \cdots \right) \,,
\label{eq:pol}
\end{equation}
where sums on repeated indices are understood.
The coefficients of the various powers of the electric field are the components of the 
susceptibility tensors. Here  $E_i$ stands for the total electric field, which could be 
in part due to the polarization itself and in part due to an applied external field. 
We have assumed that the polarization at a 
time $t$ depends only on the instantaneous value of the electric field. This is a reasonable
approximation in a regime where dispersion and dissipation can be neglected.

The wave equation for the electric field follows from Eqs.~(\ref{eq:Maxwell}) and
(\ref{eq:contitutive})  and is
 \begin{equation}
\nabla(\nabla\cdot\E) - \nabla^2\E + \frac{1}{c^2}\frac{\partial^2}{\partial t^2}\E = - \frac{1}{\epsilon_0 c^2}\frac{\partial^2}{\partial t^2}\P \,,
\label{eq:wave}
\end{equation}
where $c = 1/\sqrt(\mu_0\epsilon_0)$ is the speed of light in vacuum. We will be interested
in cases where $\nabla\cdot\E =0$, so the wave equation becomes
\begin{equation}
\biggl(\nabla^2 - \frac{1}{c^2}\frac{\partial^2}{\partial t^2}\biggr)\E =  \frac{1}{\epsilon_0 c^2}\frac{\partial^2}{\partial t^2}\P \,.
\label{eq:wave2}
\end{equation}
We will restrict our attention to solutions of this equation which are polarized in the
$z$-direction, but propagating in the $x$-direction, so that
\begin{equation}
E_i = \delta_{iz}\, E = \delta_{iz}\, E(t,x) \,.
\end{equation}
In this case, $\nabla\cdot\E =0$ is satisfied. Now we may write
\begin{equation}
P_z =  \epsilon_0 \left(\chi^{(1)} E + \chi^{(2)} E^2+ \chi^{(3)} E^3 + \cdots \right) \,,
\end{equation}
where 
\begin{equation}
\chi^{(1)} = \chi_{zz}^{(1)} \,, \quad  \chi^{(2)} =  \chi_{zzz}^{(2)} \,, \quad
 \chi^{(3)} = \chi_{zzzz}^{(3)} \,.      \label{eq:chis}
\end{equation}
Note that in introducing this notation, we are making no assumptions about isotropy.
In general the susceptibility tensors have other nonzero components which are not
equal to the components listed above. Rather, the components listed in Eq.~(\ref{eq:chis})
are the only ones needed for our discussion. 

Here we will consider nonlinear effects through third order, and ignore any higher order
effects. In this case, the wave equation for $E$ becomes
\begin{equation}
\frac{\partial^2}{\partial x^2} E -  \frac{1}{v^2}\frac{\partial^2}{\partial t^2} E 
- \frac{1}{c^2}\frac{\partial^2}{\partial t^2} \left(\chi^{(2)} E^2+ \chi^{(3)} E^3 \right) =0\,,
\label{eq:wave3}
\end{equation}
where
\begin{equation}
v = \frac{c}{\sqrt{1 + \chi^{(1)}}}
\end{equation}
is the speed of light in the $z$-direction when only the linear effects are included. 
Let $E_0(t,x)$ be a solution of Eq.~(\ref{eq:wave3}). Consider a second solution
of the form  $E = E_0 + E_1$, where $|E_1| \ll |E_0|$, but $E_1$ varies more rapidly
in time than does $E_0$, so that 
\begin{equation}
\left|\frac{\partial E_0}{\partial t}\; E_1  \right|  \ll    \left| E_0 \; \frac{\partial E_1}{\partial t}  \right| \,.
\label{eq:time_approx}
\end{equation}
Assuming $\chi^{(2)}E_1 <<1$ and $\chi^{(3)}E_0E_1 <<1$ and using the approximation
that $E_0$ is slowly varying, Eq. (\ref{eq:wave3}) reduces to the following wave equation for $E_1$,
\begin{equation}
\frac{\partial^2 E_1}{\partial x^2} - \frac{1}{v^2}\left(1 + 2 \epsilon_{1} + 3 \epsilon_{2}\right) 
\frac{\partial^2 E_{1}}{\partial t^2}  = 0,
\label{eq:we1}
\end{equation}
where
\begin{equation}
\epsilon_{1} = \frac{\chi^{(2)}}{1 + \chi^{(1)}}\,E_0(t,x),
\label{eq:eps1}
\end{equation}
and 
\begin{equation}
\epsilon_{2} = \frac{\chi^{(3)}}{1 + \chi^{(1)}}\,E_0^2(t,x) \,.
\label{eq:eps2}
\end{equation}
Note that the subscripts on $\epsilon_{1}$ and $\epsilon_{2}$ refer to the power of   $E_0$,
not the order of the susceptibility.

\subsection{WKB Solutions}
\label{sec:WKB}

Equation~(\ref{eq:we1}) describes waves which encounter a space and time dependent
effective dielectric function. In the short wavelength limit which we consider, this translates
into a propagation speed which varies in space and time. This can be made more
precise by examining WKB solutions.  We  consider plane wave  solutions of the form
\begin{equation}
E_1(t,x) = N\,{ \rm e}^{i k_1 x} \,F(t) \,,
\label{eq:sep}
\end{equation}
where $N$ is a normalization factor.  Equation~(\ref{eq:we1}) implies that
\begin{equation}
\frac{\partial^2 F}{\partial t^2} + \omega^2 F= 0 \,,
\label{eq:F}
\end{equation}
where 
\begin{equation}
\omega^2 = \omega^2(t) = \frac{\omega_1^2}{1 + 2 \epsilon_{1}(t) + 3 \epsilon_{2}(t) }\,,
\end{equation}
and $\omega_1 = k_1 v$ is the angular frequency in the limit that $\epsilon_{1} = 
\epsilon_{2} =0\,.$ A  WKB solution of Eq.~(\ref{eq:F}) is
\begin{eqnarray}
F&=& F(t,x) = (2 \omega)^{-1/2}\,\exp{\left[-i\int_0^t\,dt'\,\omega(t')\right]} \nonumber\\ 
& \approx & (2 \omega)^{-1/2}\,\exp{\left\{  -i\omega_1\,t + 
i\omega_1\int_0^t\,dt'\,\epsilon_{1}(t',x) +
 i\frac{3\,\omega_1}{2}\int_0^t\,dt'\,[\epsilon_{2}(t',x) - \epsilon^2_{1}(t',x)]  \right\}}.
\label{eq: F-soln}
\end{eqnarray}
The corresponding solution for the electric field is
\begin{equation}
E_1(t,x) = N_1\,\exp{\left\{i k_1 x - i\omega_1\,t +  i\omega_1\int_0^t\,dt'\,\epsilon_{1}(t',x) + 
i\frac{3\,\omega_1}{2}\int_0^t\,dt'\,[\epsilon_{2}(t',x)  - \epsilon^2_{1}(t',x)]        \right\}} \,,
\label{eq:E-soln}
\end{equation}
where  $N_1 = N(2 \omega)^{-1/2}$, and we are working to second order in $E_0$.

This solution describes a wave propagating with phase velocity
\begin{equation}
 u(t,x) = \frac{v}{\sqrt{1 + 2 \epsilon_{1}(t,x) + 3 \epsilon_{2}(t,x)}} 
 \approx v \,\left\{ 1 - \epsilon_{1}(t,x) - \frac{3}{2} [\epsilon_{2}(t,x)  - \epsilon^2_{1}(t,x)]   \right\}\,.
 \label{eq:u}
 \end{equation}
In order to discuss the flight times of pulses, we need to form wavepacket solutions. So long
as dispersion may be ignored, these wavepackets will have group velocities also approximately
given by Eq.~(\ref{eq:u}). Recall that this space and time dependent velocity is determined
in part by the background field, $E_0(t,x)$. This is analogous to the situation in general
relativity theory, where the apparent speed of light depends upon the spacetime geometry.
Just as in gravity, the effective lightcone for the propagation of weak disturbances depends
upon the background field.

\section{Fluctuations of the Effective Lightcone}
\label{sec:flucts}

\subsection{Flight Times of Pulses}
\label{sec:flight}

 Our analog model for quantum lightcone fluctuations arises when the background field
 $E_0(t,x)$ is no long a fixed, classical field, but is allowed to undergo quantum fluctuations.
 This leads to quantum fluctuations of the speed of light in the medium and of the effective
 lightcone. The operational meaning of these lightcone fluctuations is a variation in flight times  
of pulses between a source and a detector. If the spatial separation between the source
and detector is $r$, then  the flight time for nonzero background field  is 
 \begin{equation}
T =  \int_0^r \frac{1}{u(t(x),x)} \, dx \,.
\end{equation}
Here $x = v\, (t - t_0)$, where $t=t_0$ is the time of emission of the pulse. Hence the local 
velocity $u(t(x),x)$ is taken to be evaluated on the unperturbed path of the pulse. 
When $| \epsilon_{1}| \ll 1$ and $| \epsilon_{2}| \ll 1$, this may be written as
\begin{equation}
T \approx \frac{r}{v} +  \frac{1}{v} \,   \int_0^r 
\left[ \epsilon_{1} + \frac{1}{2} (3\epsilon_{2} - \epsilon^2_{1}) \right] \, dx \,,
\end{equation}
with $ \epsilon_{1}$ and $ \epsilon_{2}$ understood to be evaluated at  $t = t_0+x/v$.

Now we suppose that  $E_0(t,x)$, and hence $ \epsilon_{1}$ and $ \epsilon_{2}$,
undergo quantum fluctuations. The mean flight time is
\begin{equation}
\langle T  \rangle \approx \frac{r}{v} +  \frac{1}{v} \,   \int_0^r 
\left(\langle \epsilon_{1} \rangle + \frac{1}{2} \langle 3\epsilon_{2} - 
\epsilon^2_{1} \rangle \right) \, dx \,.
\end{equation}
and the variance in flight time is
\begin{equation}
(\Delta t)^2 = \langle T ^2 \rangle - \langle T  \rangle^2\,.  
\end{equation}
We will evaluate this quantity to second order in  $E_0$, which is analogous to second
order in $ h_{\mu\nu}$ in quantum gravity. This means that we retain terms of order
 $ \epsilon_{1}^2$ or $ \epsilon_{2}$, but drop higher order terms such as  
 $ \epsilon_{1} \epsilon_{2}$. The result is  
\begin{equation}
(\Delta t)^2 =  \frac{1}{v^2} \,   \int_0^r dx_1    \int_0^r dx_2 \left[ 
 \langle \epsilon_{1}(t_1,x_2)\,   \epsilon_{1}(t_2,x_2) \rangle 
-  \langle \epsilon_{1}(t_1,x_2)  \rangle   \,   \langle  \epsilon_{1}(t_2,x_2) \rangle \right]\,, 
 \label{eq:dt2-diel}
\end{equation}
where $t_i = t_0+x_i/v$. 
Note that although a term involving $\langle 3\epsilon_{2} -\epsilon^2_{1} \rangle$ appears in 
$\langle T ^2 \rangle$ and in $\langle T  \rangle^2$, it cancels from $(\Delta t)^2$.
This is analogous to the situation discussed in Sect.~\ref{sec:gravity}, where  
$ \langle \sigma_2 \rangle$ modifies the background, and hence the mean flight
time, but it is  $ \langle \sigma_1^2 \rangle$ which describes the lightcone fluctuations.

Now we assume that $ \epsilon_{1}$ fluctuates around a mean value
of zero, so that
\begin{equation}
 \langle  \epsilon_{1} \rangle = 0\,, 
\end{equation}
then the last term in Eq.~(\ref{eq:dt2-diel}) vanishes. This expression may also be written as a double integral in time: 
\begin{equation}
(\Delta t)^2 =   \int_0^{r/v} dt_1    \int_0^{r/v} dt_2 \,
\langle \epsilon_{1}(t_1)\,   \epsilon_{1}(t_2) \rangle \, , 
 \label{eq:delt2}
\end{equation}
where we drop the explicit reference to the space dependence.
Equation~(\ref{eq:delt2}) is the analog of Eq.~(\ref{eq:dt2-grav}), with the correlation
function for $ \epsilon_{1}$ playing the role of the graviton correlation function. Thus
a material with nonzero second-order susceptibility,  $ \chi^{(2)}$, is needed for 
an analog model which reproduces the features of quantum gravity described in
Sect.~\ref{sec:gravity}.  Note that if we were to work to higher order in $E_0$, then
terms including $\langle \epsilon_{2}(t_1)\,   \epsilon_{2}(t_2) \rangle$ and
$\langle \epsilon_{1}^2(t_1)\,   \epsilon_{1}^2(t_2) \rangle$ would also appear
in $(\Delta t)^2$. These terms would model the effects of quantum stress tensor fluctuations,
or passive quantum gravity effects, as opposed to the active quantum gravity fluctuations
modeled by Eq.~(\ref{eq:delt2}).

\subsection{Phase Fluctuations}
\label{sec:phase}

In addition to flight time fluctuations for wavepackets,  there is another related effect 
associated with lightcone fluctuations. This is the fluctuations in the phase of a plane 
wave solution, such as Eq.~(\ref{eq:E-soln}). Here we will assume Gaussian fluctuations,
which is the case when dealing with an approximately free quantum field. Let
\begin{equation}
\Phi = \varphi + \langle \Phi \rangle
\end{equation}
be the phase of the wave, so that $ \langle \varphi \rangle=0$. Let the fluctuations of $\varphi$
be described by a Gaussian probability distribution of width $a$ centered at  $\varphi =0$,
\begin{equation}
P(\varphi)  = \frac{1}{\sqrt{\pi}\, a} \;\exp\left(-\frac{\varphi^2}{a^2} \right) \,.
\end{equation}
The ensemble average of any function of $\varphi$ is defined by 
\begin{equation}
 \langle f(\varphi) \rangle = \int_{-\infty}^\infty  f(\varphi)\,  P(\varphi)\, d\varphi \,.
\end{equation}
One may easily verify that
\begin{equation}
 \langle {\rm e}^{i \varphi}\rangle = \exp\left( -\frac{1}{2} \,  \langle \varphi^2 \rangle \right) \,.
 \label{eq:exp}
\end{equation}

There is an alternative derivation of Eq.~(\ref{eq:exp}) in which the ensemble average is
an expectation value in a ``vacuum-like" quantum state $|\psi \rangle$, so that
$ \langle f(\varphi) \rangle = \langle \psi | f(\varphi) |\psi  \rangle$. Here  ``vacuum-like"
means that there exists a decomposition of $\varphi$, treated as a quantum operator,
into positive and negative frequency parts, $\varphi = \varphi^+  + \varphi^-$, such that
$\varphi^+ |\psi \rangle =0$.  We may write  $\langle \varphi^2 \rangle 
=  \langle \varphi^+\; \varphi^- \rangle$. This relation may be combined with   
the Campbell-Baker-Hausdorff formula, 
${\rm e}^{A + B} = {\rm e}^{-\frac{1}{2}[A,B]}\, {\rm e}^{A}\, {\rm e}^{B}$, for any pair of operators 
$A$ and $B$ that each commute with their commutator $[A,B]$ to obtain  Eq.~(\ref{eq:exp}).
The two derivations are equivalent because the  ``vacuum-like"  states are really squeezed
vacua, for which the wavefunction in a Schr{\"o}dinger representation is a Gaussian in some 
variable.

If we take the average of the electric field, as given by  Eq.~(\ref{eq:E-soln}), then the result is
\begin{equation}
\langle E_1(t,x) \rangle = A(t)\,  {\rm e}^{ i[ k_1 x - \omega_1\,t +  \phi(t) ]} \,,
\label{eq:E-ave}
\end{equation}
where
\begin{equation}
\phi(t) = \frac{3\,\omega_1}{2}\int_0^t\,dt'\,\langle\epsilon_{2}(t')  -\epsilon^2_{1}(t')  \rangle
\label{eq:phi}
\end{equation}
and
\begin{equation}
A(t) = \langle N_1 \rangle \;  {\rm e}^{ -\frac{1}{2}\omega_1^2\,  (\Delta t)^2} \, .
\label{eq:Amp}
\end{equation}
Here $(\Delta t)^2$ is given by Eq.~(\ref{eq:delt2}) with $t = r/v$.
As before, we work to second order in the background field, $E_0$, which means
second order in $\epsilon_{1}$ but first order in $\epsilon_{2}$. The effect of $\epsilon_{2}$
appears only through its expectation value, and takes the form of a phase shift which can
be interpreted as a shift in frequency of the wave. In second order, the effects of  $\epsilon_{1}$ 
also include lightcone fluctuations, which here show up as a decrease in the amplitude of the averaged wave. 

\subsection{Quantized Electric Field in a Squeezed Vacuum State}
\label{sec:sqvac}

Now we will regard the background field, $\E_0(t,\x)$, as a quantum field, which may be
expanded in terms of modes as  
\begin{equation}
 {\bf E}_0 (t,{\bf x}) = \sum_{\k \lambda} [a_{\k \lambda} {\bf \hat{e}}_{\k \lambda}  
\, f_{\k \lambda} (t,{\bf x}) + 
{a^\dagger}_{\k \lambda} {\bf \hat{e}}_{\k \lambda}  \, {f^*}_{k \lambda} (t,{\bf x}) ] \,.
\label{eq:Efield}
\end{equation}
Here $ {\bf \hat{e}}_{\k \lambda}$ are polarization vectors for linear polarization $\lambda$, 
the ${a^\dagger}_{k \lambda}$ and ${a}_{k \lambda}$
are photon creation and annihilation operators, respectively, and the mode functions are
plane waves:
\begin{equation}
 f_{\k \lambda} (t,{\bf x}) = i \sqrt{ \frac{\hbar \omega}{2\epsilon_0 \,V}}
 \, { \rm e}^{i (\k \cdot \x - \omega t)} \,.
\end{equation}
Box normalization in a volume $V$ is assumed, and $\omega = v |\k|$. 

We will take the quantum state of the electromagnetic field to be a multi-mode squeezed vacuum
state. This is analogous to the bath of gravitons discussed in Sect.~\ref{sec:gravity} and
produces lightcone fluctuations through a fluctuating effective dielectric function.
 A single mode squeezed vacuum state, $|\zeta\rangle$, is obtained by acting on the vacuum 
with the squeeze operator~\cite{Caves}
\begin{equation}
S(\zeta) = \exp{\left\{ \frac{1}{2} \left[\zeta^* \, a^2 - \zeta (a^\dagger)^2 \right]  \right\}}\, ,
\end{equation} 
where $\zeta = r \,{\rm e}^{i\eta}$ is a complex parameter, and $a^\dagger$ and $a$ are
the creation and annihilation operators for the excited mode. In this state, one has that
\begin{equation}
\langle {a}^{\dagger} {a}\rangle = \sinh^2 r \,,
\end{equation}
and
\begin{equation}
\langle  {a}^{2}\rangle = - e^{i\eta}\cosh r \sinh r \,.
\end{equation}

We now wish to construct a multi-mode squeezed vacuum state by allowing the squeeze
parameter  $\zeta$ to depend upon the mode, but assuming that
\begin{equation}
\langle  a_{\k \lambda}\,  a_{\k' \lambda'} \rangle = 
\langle  {a^\dagger}_{\k \lambda}\,  a_{\k' \lambda'} \rangle = 0
\end{equation}
unless $\k = \k'$ and $\lambda = \lambda'$. This insures that the electric field correlation 
function will be a sum of contributions from the individual modes with no cross terms.
As in Sect.~\ref{sec:wave_eq}, we wish to assume that the excited modes are all at least
approximately polarized in the $z$-direction, so that
\begin{equation}
 {\bf \hat{e}}_{\k \lambda} \approx  {\bf \hat{z}}\,,
\end{equation}
which picks out a specific polarization and allows us to omit the mode label  $\lambda$. 

Next we compute an electric field correlation function as an expectation value of a product
of $z$-components of
electric field operators. We are interested in the effects of excited modes, and take the
product to be normal ordered and symmetrized to write 
\begin{eqnarray}
\langle E_0(t,\x)\,  E_0(t',\x') \rangle &=& 
\frac{1}{2} \langle :E^z_0(t,\x)\,  E^z_0(t',\x'): +  :E^z_0(t',\x')\,  E^z_0(t,\x): \rangle \\ \nonumber
&=& \frac{\hbar}{\epsilon_0\,V}\,  \textrm{Re}\sum_{\k} \omega
\left\{ \langle {a_\k}^{\dagger} {a_\k}\rangle \, { \rm e}^{i [\k \cdot (\x -\x')- \omega (t-t')]}
-\langle {a_\k}^2 \rangle \, { \rm e}^{i[ \k \cdot (\x +\x')- \omega (t+t')]} \right\} \,.
\end{eqnarray}
When each of the excited modes is a squeezed vacuum state, this becomes
\begin{eqnarray}
\langle E_0(t,\x)\,  E_0(t',\x') \rangle &=& \frac{\hbar}{\epsilon_0\,V}\,  \sum_{\k} \omega \,
\sinh r_\k\, \left\{ \sinh r_\k\, \cos[ \k \cdot (\x -\x')- \omega (t-t')]   \right.  \nonumber \\
&+&  \left. \cosh r_\k\, \cos[\k \cdot (\x +\x')- \omega (t+t') +\eta_\k] \right\} \,.
\label{eq:E-corr}
\end{eqnarray}
The mean squared electric field is the coincidence limit of this expression
\begin{equation}
\langle E_0^2(t,\x) \rangle =  \frac{\hbar}{\epsilon_0\,V}\,  \sum_{\k} \omega \,
\sinh r_\k\, \left\{ \sinh r_\k +  \cosh r_\k\, \cos[2\k \cdot \x- 2\omega t +\eta_\k] \right\} \,.
\label{eq:E2}
\end{equation}
Note that this quantity can be negative, which is an example of a subvacuum effect,
similar to negative energy density in quantum field theory~\cite{FR11}. 

The phase shift, $\phi$, defined in Eq.~(\ref{eq:phi}), can be expressed as a time
integral of the squared electric field as
\begin{equation}
\phi(t) = \frac{3\,\omega_1}{2(1 + \chi^{(1)})} \,
\left[\chi^{(3)} - \frac{(\chi^{(2)})^2  }{1 + \chi^{(1)}} \right]
 \; \int_0^t\,dt'\, \langle E_0^2(t',\x') \rangle \,.
\end{equation}
Similarly, the mean squared time delay, given by Eq.~(\ref{eq:delt2}), may be written as
 \begin{equation}
(\Delta t)^2 = \left( \frac{\chi^{(2)}}{1 + \chi^{(1)}}\right)^2   \, \int_0^{t} dt'    \int_0^{t} dt'' \,
\langle E_0(t',\x')\,  E_0(t'',\x'') \rangle \,.
\end{equation}
Here we understand $\x'(t')$ and  $\x''(t'')$ to be on the mean path of the light ray.
We assume that the excited modes are all approximately moving in the $+x$-direction,
so that $\x(t) \approx v\, t  \mathbf{\hat{x}}$. However, in the limit of large quantization 
volume $V$, there are necessarily still a large number of excited modes, and we may
write
\begin{equation}
\frac{1}{V} \sum_{\k} \rightarrow  \frac{1}{(2\pi)^3} \int d^3 k \approx
 \frac{1}{(2\pi)^2}\, k^2\,\Delta k \, \Delta \theta \,.   
 \label{eq:bandwidth}
\end{equation}
Here $\Delta k \ll k$ is the bandwidth in wavenumber of the excited modes, and 
$\Delta \theta \ll 1$ is their angular spread around the  $+x$-direction. As in
Sect.~\ref{sec:wave_eq}, we assume that all of the relevant modes have a phase 
velocity of $v =\omega/k$, so that
\begin{equation}
 \k \cdot \x \approx \omega t \,.
 \label{eq:x-dir}
\end{equation}
This causes the arguments of the trigonometric functions in Eqs.~\eqref{eq:E-corr} and
\eqref{eq:E2} to be approximately constant. 
Here the electric field correlation function and squared
electric field become time-independent, and given by
\begin{equation}
\langle E_0(t,\x)\,  E_0(t',\x') \rangle = \langle E_0^2 \rangle = 
 \frac{\hbar}{4 \pi^2\, \epsilon_0}\,   \omega \, k^2\,\Delta k \, \Delta \theta \,   
\sinh r  ( \sinh r+  \cosh r \cos \eta ) \,.
\end{equation}
Here we have used Eq.~\eqref{eq:bandwidth} and assumed that the squeeze parameter
is approximately the same for all excited modes, so the mode label on $r$ and $\eta$
has been dropped. 

Note that the energy density $U$ in empty space associated with the excited modes is
\begin{equation}
U = \epsilon_0 \,  \langle E_0^2 \rangle \,.
\end{equation}
It will be convenient to express our final results in terms of $U$. The phase shift $\phi$
becomes
\begin{equation}
\phi(t) = \frac{U}{2\epsilon_0}\frac{3\,t}{1 + \chi^{(1)}} \,
\left[\chi^{(3)} - \frac{(\chi^{(2)})^2}{1 + \chi^{(1)}} \right] \,.
\label{eq:phi2}
\end{equation}
Because $\phi \propto t$, it can be interpreted as a shift in frequency of the original
probe field of
\begin{equation}
\delta \omega = -  \frac{3 U}{2\epsilon_0 (1 + \chi^{(1)})} \,
\left[\chi^{(3)} - \frac{(\chi^{(2)})^2}{1 + \chi^{(1)}} \right]\,.
\end{equation}
More interesting for our purposes is the variance in flight times of pulses, which becomes
\begin{equation}
(\Delta t)^2 =  \frac{U}{\epsilon_0}\biggl(\frac{\,\chi^{(2)}}{1 + \chi^{(1)}}\biggr)^2\;  t^2   \,.
\end{equation}
This result is the analog of Eq.~\eqref{eq:sq-grav}, and   tells us that the root-mean-squared 
fluctuation in flight time can grow linearly with increasing
path length. Thus the lightcone fluctuations in a nonlinear material with nonzero $\chi^{(2)}$
can potentially become large. The correlation time in this case is expected to be of the order
of the period of the excited modes in the squeezed state, which form the background field.
Thus pulses separated by more than this period should have flight time variations of order
$\Delta t$.

\section{Some Numerical Estimates}
\label{sec:est}

Here we wish to give an estimate of the magnitude of the lightcone fluctuation effect in
nonlinear materials and an assessment of the possibility of observing it in an experiment.
First, let the ratio of the root-mean-squared fluctuation in flight time, 
$\Delta t_{\rm rms} = \sqrt{(\Delta t)^2}$, to the total flight time be denoted by $\delta$, so
that 
\begin{equation}
\delta = \frac{\Delta t_{\rm rms}}{t} =  \sqrt{ \frac{U}{\epsilon_0}}\, \frac{\chi^{(2)}}{1 + \chi^{(1)}}\,.
\label{eq:delta}
\end{equation}
This may be expressed as
\begin{equation}
\delta = \frac{3.36 \times 10^{-6}}{1 + \chi^{(1)}} \; \frac{\,\chi^{(2)}}{10^{-11}\, {\rm m/V}} \; 
 \sqrt{ \frac{U}{ 1 \,{\rm J/m^3}}} \,.
 \label{eq:delta2}
\end{equation}

We need to make estimates of each of the parameters which appear in Eq.~\eqref{eq:delta2}.
The first order susceptibility, $ \chi^{(1)}$ is typically of order or less than unity at optical frequencies,
so the $1/(1 + \chi^{(1)})$ factor will be of order one. The second order susceptibility, $ \chi^{(2)}$ is 
more variable, and can only be nonzero for materials which do not possess spatial inversion
symmetry (i.e., are noncentrosymmetric). Such materials typically exhibit a significant degree of
anisotropy, so the different components of the tensor $ \chi_{ijk}^{(2)}$ vary considerably in
magnitude. We need a material for which a diagonal component ($i=j=k$) of this tensor is 
nonzero, which we can take to be  $ \chi_{zzz}^{(2)} =   \chi^{(2)}$. An example of such a
material is cadmium sulfide (CdS), for which $ \chi^{(2)} \approx 4 \times 10^{-11}\,  {\rm m/V}$.
(See Fig.~22 in Ref.~\cite{Shoji}, noting~\cite{Boyd} that $\chi_{zzz}^{(2)} = 2 d_{33}$.)
Other materials, such as gallium arsenide, possess somewhat higher values of the
second order susceptibility, but have only off-diagonal components of $ \chi_{ijk}^{(2)}$ being
nonzero~\cite{Boyd}.

Next we need an estimate of the realizable energy density in a squeezed vacuum state.
Experiments producing squeezed light have achieved $10\,{\rm db}$ of squeezing~\cite{sq2}, 
meaning 
$10 = 10\log_{10}\bigl(U/U_v\bigr)$, where $U_v$ is the vacuum energy density,  so that 
$U \approx 10\,U_v$ for each squeezed mode, or about five photons per mode. The number
of modes excited in a finite quantization volume $V$ is the product of $V$ with the right hand
side of Eq.~\eqref{eq:bandwidth}. Thus
\begin{equation}
U \approx  \frac{5 \hbar \omega}{(2\pi)^2}\, k^2\,\Delta k \, \Delta \theta 
=  20 \pi^2\, \frac{\hbar c}{\lambda^4} \, \left(\frac{\Delta k}{k}\right) \, \Delta \theta   \,,  
\end{equation}
where $\lambda = 2 \pi/k$ is the mean wavelength of the excited modes, which will
typically be in the infrared or optical ranges. This can be
written as
\begin{equation}
U \approx 6.25\, {\rm \frac{J}{m^3}} \, \left(\frac{1000\, {\rm nm}}{\lambda}\right)^4\, 
 \left(\frac{\Delta k}{k}\right) \, \Delta \theta   \,.
\end{equation}
To be consistent with our earlier analysis, we need both  $\Delta k/k$ and  $\Delta \theta$  
to be small. If, for example, we have  $\Delta k/k   \approx \Delta \theta \approx 10^{-2}$, then
fractional flight time variations of order $\delta \approx 10^{-9}$ would seem to be possible using
infrared light. Although this is small, it might be observable with accurate flight time 
measurements.

Recall from Eq.~\eqref{eq:Amp}, that the growth of $(\Delta t)^2$ with flight time causes
the amplitude of a plane wave solution to decay exponentially in time. This is a feature
of dephasing, where phase fluctuations lead, via Eq.~\eqref{eq:exp}, to a decaying amplitude.
This is also in principle an observable characteristic of lightcone fluctuations. However in
an experiment, it may be very difficult to distinguish this effect from the exponential decay
due to other forms of absorption of the wave.

Although gravity is dispersionless, real dielectric materials exhibit frequency dependence
in their susceptibilities, which we have ignored. In addition, we have assumed that the
frequency of the probe field is higher than that of the background field. In general, the
two fields will experience different values of the first and second order susceptibilities,
which we have not considered. However, many materials with nonzero   $\chi^{(2)}$,
including CdS, exhibit a $\chi^{(2)}(\omega)$ which is relatively independent of frequency
in the infrared and the red end of the visible spectrum~\cite{Shoji,Wagner}. The second
order susceptibility starts to vary more strongly with frequency at higher frequencies. 
Thus our analysis should give an approximate description of lightcone fluctuation effects
in the red and infrared parts of the spectrum. There has been recent progress on designing
quantum dot hetereostructures with specified second order susceptibilities~\cite{Zielinski}.
This raises the possibility of engineering materials with the properties needed for this
type of experiment.

\section{Summary and Discussion}
\label{sec:sum}

In this paper, we have argued that nonlinear optical effects provide an analog model for
quantum lightcone fluctuations in quantum gravity. The basic idea is that a fluctuating background
electric field in a nonlinear material produces a fluctuating effective dielectric function for
a probe field, leading to fluctuations in the effective speed of light. This phenomenon is not
unique to nonlinear optics, but will arise in any system with a nonlinear wave equation. Sound
wave in a fluid, for example, satisfy the nonlinear Westerwelt equation. Thus a fluctuating
background sound wave can induce fluctuation of the ``soundcone", and hence leading
to a similar analog model. However, nonlinear optics seems to be a more promising arena
for precision measurements. 

Materials with a nonzero second order susceptibility, $\chi^{(2)}$, provide the best analog
models for the effects in linearized quantum gravity. The variance of the flight time of a pulse,
which can be taken to be the signature of lightcone fluctuations, is proportional to an integral
of an electric field correlation function in the analog model, while the same quantity is an
integral of the graviton correlation function in quantum gravity.  We considered a background
electric field created by photons in a multimode squeezed vacuum state, which induces
quantum fluctuations in the effective dielectric function experienced by probe pulses. Our
numerical estimates suggest that fractional flight time variations of order $10^{-9}$ might be
achievable and potentially measurable. An alternative to photons in squeezed vacuum states
might be a stochastic bath of electromagnetic waves, for which $\langle E_0 \rangle =0$.
This might allow even larger effects. 

Given that quantum gravity itself is generally remote from experimental tests, further study
of analog modes such as that proposed in this paper seems to be a promising area for
further enquiry.

\begin{acknowledgments}
This work was supported in part  by the National Science Foundation under
Grant PHY-0855360, by Funda\c{c}\~ao de Amparo a 
Pesquisa de S\~ao Paulo (FAPESP), Funda\c{c}\~ao de Amparo \`a Pesquisa do Estado 
de Minas Gerais (FAPEMIG), and Conselho Nacional de Desenvolvimento Cient\'\i fico 
e Tecnol\'ogico do Brasil (CNPq). 

\end{acknowledgments}


\begin{thebibliography}{99}

\bibitem{Pauli} W. Pauli, Helv. Phys. Acta. Suppl. {\bf 4}, 69 (1956).
This reference consists of some
remarks made by Pauli during the discussion of a talk by O. Klein at a 1955
conference in Bern, on the 50th anniversary of relativity theory.

\bibitem{Deser}  S. Deser, Rev. Mod Phys. {\bf 29}, 417 (1957).

\bibitem{DeWitt}  B. S. DeWitt, Phys. Rev. Lett. {\bf 13}, 114 (1964).

\bibitem{Wheeler} J.A. Wheeler, {\it Geometrodynamics}, (Academic Press, New York, 1963).

\bibitem{Hawking} S.W. Hawking, Phys. Rev. D  {\bf 37}, 904 (1988). 

\bibitem{Ng} W.A. Christiansen, Y.J. Ng, D.J.E. Floyd,
and E.S. Perlman, Phys. Rev. D {\bf 83}, 084003 (2011), arXiv:0912.0535.

\bibitem{EMN00} J. Ellis, N.E. Mavromatos, and D.V. Nanopoulos, Phys. Rev. D  {\bf 61},
027503 (1999), arXiv:gr-qc/9906029. 

\bibitem{F95}  L. H. Ford, Phys. Rev. D {\bf 51}, 1692 (1995), arXiv:gr-qc/9410047.

\bibitem{FS96} L. H. Ford and N. F. Svaiter,  Phys. Rev. D {\bf 54}, 2640 (1996),
arXiv:gr-qc/9604052. 

\bibitem{FS97}  L. H. Ford and N. F. Svaiter,  Phys. Rev. D  {\bf 56}, 2226 (1997),
arXiv:gr-qc/9704050.

\bibitem{FW08} For a review, see for example, L.H. Ford and C-H Wu, AIP Conf. Proc.
{\bf 977} 145 (2008),  arXiv:0710.3787.

\bibitem{BLV} C. Barcelo, S. Liberati, and M. Visser, Living Rev. Rel. {\bf 8}, 12 (2005),
arXiv:gr-qc/0505065.  

\bibitem{KMN10} G. Krein, G. Menezes and N. F. Svaiter, Phys. Rev. Lett. {\bf 105}, 131301 (2010).

\bibitem{AGKMS11} E. Arias, E. Goulart, G. Krein, G. Menezes, and N. F. Svaiter,
Phys. Rev. D {\bf 83}, 125022 (2011).

\bibitem{Gordon} W. Gordon, Ann. Phys. (Leipzig) {\bf 72}, 421 (1923).

\bibitem{Moller} See, for example, C. M{\o}ller, {\it The Theory of Relativity}, (Oxford University Press,
London, UK, 1952 ),  pp 302-309.


\bibitem{Shapiro} I.I. Shapiro,   Phys. Rev. Lett. {\bf 13}, 789 (1964).

\bibitem{Will} C.M. Will, Annalen Phys. {\bf 15}, 19 (2005), arXiv:gr-qc/0504086.

\bibitem{YF99}   H. Yu and L.H. Ford,   Phys. Rev. D  {\bf 60}, 084023 (1999),
arXiv:gr-qc/9904082.

\bibitem{YSF09} H. Yu, N.F. Svaiter, and L.H. Ford,   Phys. Rev. D  {\bf 80}, 124019 (2009),
arXiv:0904.1087. 

 \bibitem{BF04} J. Borgman and L.H. Ford,   Phys. Rev. D  {\bf 70}, 064032 (2004),
 arXiv:gr-qc/0307043.
 
\bibitem{TF06} R.T. Thompson  and L.H. Ford,   Phys. Rev. D  {\bf 74}, 024012 (2006),
arXiv:gr-qc/0601137.    

\bibitem{Boyd} Robert W. Boyd, {\it Nonlinear optics}, 3rd ed. (Academic Press, New York, 2008).

\bibitem{Caves} C. M. Caves, Phys. Rev. D {\bf 23}, 1693 (1981).

\bibitem{FR11}   L. H. Ford and T.A. Roman, Annals  Phys. {\bf 326}, 2294 (2011),
arXiv:0907.1638. 

\bibitem{Shoji} I. Shoji, T. Kondo, A. Kitamoto,  M. Shirane, and R. Ito, J. Opt. Soc. Am. B
{\bf 14},  2268 (1997).

\bibitem{sq2} H. Vahlbruch, M. Mehmet, S. Chelkowski, B. Hage, A. Franzen, N. Lastzka, 
S. Go{\ss}ler, K. Danzmann, and R. Schnabel, Phys. Rev. Lett. {\bf 100}, 033602 (2008).

\bibitem{Wagner} H.P. Wagner, M. K{\"u}hnelt, W. Langbein, and J.M. Hvam,
Phys. Rev. B {\bf 58}, 10494 (1998).  

\bibitem{Zielinski} M. Zielinski, S. Winter, R. Kolkowski, C. Nogues, D. Oron, J. Zyss, and 
D. Chauvet, Optics Express {\bf 19} 6657 (2011).



\end{thebibliography}
\end{document}